\documentclass[tightenlines, reprint, aps, prb]{revtex4-2}

\usepackage{amsmath,amsfonts,amssymb,mathrsfs}
\numberwithin{equation}{section}
\renewcommand\theequation{\arabic{section}.\arabic{equation}}
\usepackage{color,graphicx}
\usepackage{hyperref}
\hypersetup{
    colorlinks=true,
    linkcolor=blue,
    filecolor=magenta,      
    urlcolor=blue,
    citecolor=magenta,
}

\usepackage{color}

\newcommand{\B}{\mathbf}
\let\oldref\ref
\renewcommand{\ref}[1]{(\oldref{#1})}

\newcommand\numberthis{\addtocounter{equation}{1}\tag{\theequation}}

\begin{document}
\title{\vspace{-3.0cm}\textbf{The functional-integral approach to Gaussian fluctuations in Eliashberg theory}}

\author{Mason Protter, Rufus Boyack, \& Frank Marsiglio}
\affiliation{Department of Physics \& Theoretical Physics Institute, University of Alberta, Edmonton, Alberta T6G~2E1, Canada}

\begin{abstract}
The Eliashberg theory of superconductivity is based on a dynamical electron-phonon interaction as opposed to a static interaction present in BCS theory. 
The standard derivation of Eliashberg theory is based on an equation of motion approach, which incorporates certain approximations such as Migdal's approximation for the pairing vertex. 
In this paper we provide a functional-integral-based derivation of Eliashberg theory and we also consider its Gaussian-fluctuation extension. 
The functional approach enables a self-consistent method of computing the mean-field equations, which arise as saddle-point conditions, and here we observe 
that the conventional Eliashberg self energy and pairing function both appear as Hubbard-Stratonovich transformations. 
An important consequence of this fact is that it provides a systematic derivation of the Cooper and density-channel interactions in the Gaussian fluctuation response. 
We also investigate the strong-coupling fluctuation diamagnetic susceptibility near the critical temperature. 
\end{abstract}

\maketitle

\small

\section{Introduction}

In conventional superconductivity, the ordered phase originates from an effective attraction between electrons driving the formation of Cooper pairs. Eliashberg theory~\cite{Eliashberg1960,Eliashberg1961} is a framework where the effective electron attraction arises from a dynamical electron-phonon interaction. Unlike BCS theory,  which is based on a static attractive interaction, Eliashberg theory has a frequency-dependent pairing function. Early reviews on Eliashberg theory include Refs.~\onlinecite{Parks1,Parks2,Bardeen1973}, a subsequent survey of the calculation of the critical temperature in Ref.~\onlinecite{Allen1982}, and more recently Refs.~\onlinecite{Bennemann,Marsiglio2020} have given up-to-date accounts on the subject matter. One of the standard approaches to deriving the self-consistent Eliashberg equations involves the equation of motion method.~\cite{Rickayzen,Marsiglio2020} As outlined in Ref.~\onlinecite{Chubukov2020}, there are a number of approximations in the standard Eliashberg theory, such as Migdal's approximation~\cite{Migdal1958} and ignoring particle-hole interactions.~\cite{Phan2020} While these assumptions can be tested a posteriori, it is beneficial to have a robust theoretical treatment where the approximations made at each stage of the derivation are clearly articulated. 

There has recently~\cite{Liu2019,Liu2020} been a new approach to studying systems with electron-phonon interactions. The method is again based on an equation of motion approach, but now it is cast in the form of the functional Schwinger-Dyson equations.~\cite{RyderBook} One advantage of this formulation of the problem is that it allows the tools from quantum field theory to be utilised and in particular the Ward identities for the electron-phonon system can be derived in the same manner as the Ward identities for quantum electrodynamics. Other functional-integral applications to electron-phonon systems include topological superfluids~\cite{Wu2016} and generalizations of the Sachdev-Ye-Kitaev model.~\cite{Esterlis2019, Hauck2020}
With this renewed interest in functional formulations of Eliashberg theory, here we aim to understand electron-phonon systems by using a functional-integral approach, and in addition we aim to bring the theoretical understanding of the Eliashberg fluctuation theory to the same level as the well-understood BCS counterpart.~\cite{VarlamovBook,AltlandSimons,Taylor2006}

In this paper, we develop a model describing electron-phonon interactions and we derive the mean-field Eliashberg equations using a functional-integral approach, which achieves what has been done~\cite{VarlamovBook,AltlandSimons,Taylor2006} for the counterpart BCS theory. 
We show that when the effective electronic interactions are decoupled in the density and Cooper channels, the saddle-point solutions of this description correspond to the Eliashberg theory. 
Deriving Eliashberg theory from the functional-integral formalism has the benefit of providing an alternative way to understand the theory compared to the canonical derivation, 
making it accessible to a wider range of physicists and also enabling those familiar with the canonical derivations to see the theory in a new light. 

A natural extension of the functional formalism is to consider fluctuations of the HS fields about their mean-field values. 
In the case of conventional Gaussian-fluctuation theory, normal-state fluctuations have been well studied in the diagrammatic~\cite{VarlamovBook,Varlamov2018} and functional-integral~\cite{Svidzinskii1971,SvidzinskiiBook} approaches, 
and similarly the extension to fluctuations in the superconducting phase have also been derived using the functional approach.~\cite{SvidzinskiiBook, Svidzinskii1971,Taylor2006,Anderson2016} 
The advantage of the functional approach lies in manifestly ensuring that gauge invariance and thermodynamic sum rules are satisfied.~\cite{Anderson2016} 

Normal-state fluctuations within Eliashberg theory have not been extensively investigated, and the main focus has been on the diagrammatic calculation of the Aslamazov-Larkin contribution to the fluctuation conductivity,~\cite{Narozhny1993} 
the specific-heat correction,~\cite{Narozhny1994} and more recently the Hall and Nernst effects.~\cite{Levchenko2020}
The functional-integral formalism we develop enables us to systematically derive the fluctuation action and response at the Gaussian level, and moreover, 
since this method clearly identifies the apposite fields which can fluctuate, we will rigorously derive the physics investigated in Refs.~\onlinecite{Narozhny1993,Narozhny1994}. 
Our paper thus provides the theoretical framework for the fluctuation formalism for Eliashberg superconductors.
An experimental quantity of interest~\cite{Ong2010} is the diamagnetic susceptibility, and here we calculate the fluctuation contribution in the strong-coupling limit. 

The structure of the paper is as follows. In Sec.~\ref{sec:EPh} we present the electron-phonon model and exactly integrate out the phonons to obtain an attractive interaction. 
In Sec.~\ref{sec:Supcon} we perform the Hubbard-Stratonovich analysis and then in Sec.~\ref{sec:HST} we obtain the mean-field Eliashberg equations. 
The Gaussian fluctuation response is studied in Sec.~\ref{sec:FlucRes} and in particular we compute the Aslamazov-Larkin contribution to the diamagnetic susceptibility near the critical temperature. 
The conclusion is presented in Sec.~\ref{sec:Conc}.

\section{The electron-phonon interaction}
\label{sec:EPh}
\subsection{The Model}

We consider the following model Hamiltonian incorporating electron-phonon interactions:~\cite{Rickayzen,AltlandSimons}
\begin{align}\label{eq:Hblf}
  \hat{H} & = \sum_{\B{k},\sigma} \epsilon_{\B{k}}  \hat{c}^\dagger_{\B{k}  \sigma} \hat{c}_{\B{k}  \sigma}+\sum_{\B{q}} \Omega_{\B{q}} \hat{b}^\dagger_{\B{q}} \hat{b}_{\B{q}}\nonumber\\
              &\quad+ \sum_{\B{k,q},\sigma}g_{\B{q}}\left(\hat{b}_{\B{q}}+ \hat{b}^\dagger_{-{\B{q}}}\right)\left(\hat{c}^\dagger_{\B{k} \uparrow} \hat{c}_{\B{k+q} \uparrow}+\hat{c}^\dagger_{\B{-k-q} \downarrow} \hat{c}_{\B{-k} \downarrow}\right). \numberthis
\end{align}
Here, $\hat{c}^\dagger$ and $\hat{b}^\dagger$ are the respective electron and phonon creation operators, whereas $\hat{c}$ and $\hat{b}$ are the corresponding annihilation operators.  The electron and phonon dispersion relations are respectively denoted by $\epsilon_{\B k}$ and $\Omega_{\B{q}}$. The electron-phonon interaction is $g_{\B{q}}$, which satisfies $g^*_{\B{q}}=g_{-\B{q}}$. In the case of a multi-branch phonon, we take $\B{q}$ to also include branch indices. 

As in BCS theory, this Hamiltonian describes the scattering of pairs of electrons with equal and opposite momentum and spin.~\cite{Rickayzen}
To ensure this type of scattering is predominant, a macroscopic number of phonons with equal momentum are introduced. 
The interaction physically describes the scattering of an electron from $\B{k+q}\uparrow$ to $\B{k}\uparrow$ due to the emission of a phonon of momentum $\B{q}$, 
upon which the phonon can be absorbed by scattering an electron from $\B{-k-q}\downarrow$ to $\B{-k}\downarrow$. 

To examine the model in Eq.~\eqref{eq:Hblf} using the finite-temperature functional-integral formalism,~\cite{AltlandSimons} we construct an action functional through the Legendre transformation:
\begin{align}
\label{eq:action1}
  S[\bar c, c, \bar b, b] = \int_0^\beta d\tau \bigg[\sum_{\B{k},\sigma}
  &\bar c_{\B{k},\sigma} \left(\partial_\tau + \mu\right) c_{\B k} +  \sum_{\B q}\bar b_{\B q} \partial_\tau b_{\B q}\nonumber\\
  &\quad + H(\bar c, c, \bar b, b)\bigg]
\end{align}
Here, $\tau$ denotes imaginary time and $\beta=1/T$. In this paper we use Natural units $c=\hbar=k_{B}=1$.
The operators $\hat{c}^\dagger,\,\hat{c}, \, \hat{b}^\dagger, \, \hat{b}$ from the Hamiltonian formalism have been replaced by the Grassmann-valued
functions $\bar c(\tau),~ c(\tau)$ and the complex-valued functions $\bar b(\tau),~b(\tau)$ respectively. 
The fermionic chemical potential is denoted by $\mu$, whereas since the number of phonons is not fixed the phonons have zero chemical potential.~\cite{AGD}

We can express the field operators in terms of Matsubara frequencies by taking the Fourier transform of the imaginary-time expressions:
\begin{equation}
  c(\tau) = {1 \over {\beta}}\sum_{\omega_n} c_n e^{-i \omega_n \tau}; \quad
  b(\tau) = {1 \over {\beta}} \sum_{\Omega_m} b_m e^{-i \Omega_m \tau}. 
\end{equation}
Here, $\omega_n \equiv (2n + 1)\pi  T$, where $n\in\mathbb{Z}$, are the fermionic Matsubara frequencies and $\Omega_m \equiv 2m \pi T$, where $m\in\mathbb{Z}$, are the bosonic Matsubara frequencies. 
Defining the four-vector momenta by $k \equiv (i\omega_n, \B k)$ and $q \equiv (i\Omega_m, \B q)$, the action in Eq.~\eqref{eq:action1} becomes
\begin{align}
  \label{eq:action}
    S[\bar c, c, \bar b, b] &= -\sum_{k,\sigma}\bar c_{k \sigma} G_0^{-1}(k) c_{k \sigma}  -\sum_{q} \Bar{b}_q D^{-1}_0(q) b_q \nonumber\\
&\quad+\sum_{k,q,\sigma} g_{\B q}\left(b_q + \Bar b_{-q} \right) \bar{c}_{k\sigma} c_{k+q \sigma}.
\end{align}
The free-particle fermionic (inverse) Green's function is defined by $ G_0^{-1}(k) \equiv i \omega_n - \xi_{\B k}$, where $\xi_{\B k}=\epsilon_{\B k}-\mu$, and the counterpart for phonons is defined by  $D^{-1}_0(q) \equiv i \Omega_m - \Omega_{\B q}$. 
Given this action functional, the finite-temperature partition function is then $Z = \int \mathfrak{D}[\bar c, c, \bar b, b] e^{-S[\bar c,\, c,\, \bar b,\, b]}$.

\subsection{Integrating out the phonons}

The fermionic density function is defined as $\rho_{q} \equiv \sum_{k, \sigma}\bar c_{k, \sigma} c_{k+q, \sigma}$ and it satisfies $\rho_{q}=\bar{\rho}_{-q}$. 
Using this definition, we can factor out the part of the partition function depending on the phonon variables and write
\begin{equation}
\label{eq:Sph1}
Z_{ph}[\bar c, c]= \int \mathfrak{D}[\bar{b},b]e^{\sum_{q}\left[\bar{b}_{q} D^{-1}_0 b_{q} - g_{\B q}\left(b_{q} + \bar{b}_{-q} \right) \rho_{q}\right]}.
\end{equation}
The functional integral over the fields $\bar{b}$ and $b$ is Gaussian, and thus it can be computed exactly by performing shifts in the fields given by $\bar{b}_{q} \rightarrow \bar{b}_{q} + g_{\B q} D_0(q)\bar{\rho}_{-q} $ and $b_{q} \rightarrow b_{q} + g_{-\B q} D_{0}(q)\rho_{-q}$.
As in the previous section, $\bar{b}_{q}\neq b_{-q}$ and thus we can perform independent transformations of these fields. 
Upon performing these transformations, $S_{ph}$ is then written explicitly as a quadratic function of $\bar b$ and $b$, and thus the functional integral may be computed exactly using the standard formula.~\cite{AltlandSimons}
The result is
\begin{align}
  Z_{ph}[\bar c, c]
  &= \int \mathfrak{D}[\bar{b},~b]e^{\sum_{q}\left(\bar{b}_qD^{-1}_{0}(q)b_q  -  |g_{\B q}|^2 D_0(q)\rho_q\bar{\rho}_q\right) } \nonumber\\
  &=\mathcal{N} e^{-\sum_{q} {|g_{\B q}|^2} D_0(q) \rho_q \bar\rho_q}. 
\end{align}
The prefactor $\mathcal{N}$ is an unimportant constant independent of $\rho_{q}$. In obtaining the result above, we have used the properties $g^{*}_{\B q}=g_{-\B q}$ and $\bar{\rho}_{q}=\rho_{-q}$. 
While $D_{0}(q)$ is the non-interacting phonon propagator, notice that in the effective interaction term this propagator is coupled to a quantity that is manifestly even in $q$. 
Therefore, only the even part of $D_{0}(q)$ contributes to the sum, i.e., only the propagator for the \emph{real} part of the phonon field, $b_q + \bar{b}_{-q}$, is important.
Hence, we write the phonon part of the partition function as
\begin{equation}
  Z_{ph}[\bar{c},c] \sim e^{-{1 \over 2} \sum_{q} {|g_{\B{q}}|^2} \mathcal{D}_0(q) \rho_{q} \bar\rho_q}, 
\end{equation}
where the phonon propagator $\mathcal{D}_{0}$ appearing above is:~\cite{AGD}
\begin{equation}
  \mathcal{D}_0(q) \equiv D_0(q) + D_0(-q) = {2 \Omega_q \over (i\Omega_m) ^2 - \Omega_q^2}.
\end{equation}
The propagator $\mathcal{D}_{0}$ describes longitudinal excitations in an isotropic medium, and necessarily it is a real quantity since it corresponds to real displacements of atoms on a lattice. 
Moreover, this property holds even in the presence of interactions. The full partition function thus reduces to 
\begin{equation}
  Z = \int \mathfrak{D}[\bar c, c]  \exp\left[\sum_{k, \sigma} \bar c_{k \sigma}G_0^{-1}(k)c_{k \sigma}
  -{1 \over 2} \sum_q \lambda_q \rho_q \bar \rho_q \right],
\end{equation}
where $\lambda(q) \equiv |g_{\B q}|^2  \mathcal{D}(q)$ 
and we have ignored the prefactor term as it does not affect the electronic quantities of interest. 
At this stage of the development, integrating out the phonons has resulted in a dynamical density-density interaction term for the fermion fields and the full partition function is of the form
$\int \mathfrak{D}[\bar c, c] e^{-S_F[\bar c,~ c]}$. The dynamical properties of this interaction term give rise to a frequency-dependent pairing function, which does not occur in the BCS theory of superconductivity. 
Formulated in real space, the fermionic action reads
\begin{align} 
\label{eq:SF}
S_F &=  \sum_{x,y, \sigma} \bar c_{x \sigma} \, G_{0}^{-1}(x-y)\, c_{y \sigma}\nonumber\\
&\quad + {1 \over 2}\sum_{x,y,\sigma,\sigma^\prime} \lambda(x-y) \bar{c}_{x\sigma} c_{x\sigma} \bar{c}_{y\sigma^{\prime}} c_{y\sigma^{\prime}}~. 
\end{align}
Sums over repeated spin labels are taken to be implicit from now on.

\section{Superconductivity in the Cooper Channel}
\label{sec:Supcon}

\subsection{Hubbard Stratonovich transformation}

The interaction term in the fermionic action in Eq.~\eqref{eq:SF} is  quartic in the fermionic fields $\bar c$ and $c$. 
At present there are no known techniques for exactly computing the functional integral of such an interaction, 
but we can make progress by reformulating the calculation in terms of \emph{auxillary fields} using the Hubbard-Stratonovitch transformation. 
This procedure eliminates the interaction term at the cost of introducing a functional integral over the axuiliary fields.
We first expand the interaction term in the action into the two parts where $\sigma = \sigma'$ and $\sigma \neq \sigma'$, which results in
\begin{align}
S_I[\bar c,c]&\equiv {1 \over 2} \sum_{x,y} \lambda(x-y)\bar{c}_{x,\sigma} c_{x,\sigma} \bar{c}_{y,\sigma^{\prime}} c_{y,\sigma^{\prime}} \nonumber\\
                   & = \sum_{x,y}\lambda(x-y)\Big({1 \over 2} \bar{c}_{x\sigma} c_{x\sigma} \bar{c}_{y\sigma} c_{y\sigma}
                          + \bar{c}_{x\uparrow}  \bar{c}_{y\downarrow} c_{y \downarrow} c_{x \uparrow}\Big).
\end{align}
We now introduce the bosonic auxillary fields $\bar\phi$, $\phi$, $\Sigma$ and a measure $\mathfrak{D}[\bar\phi,\phi,\Sigma]$ chosen such that
 \begin{equation}
   1 = \int \mathfrak{D}[\bar\phi,\phi,\Sigma]\exp\left[
   -\sum_{x,y}{\bar\phi_{xy} \phi_{xy} + {1 \over 2}\Sigma_{xy}^{\sigma}\Sigma_{yx}^{\sigma} \over \lambda(x-y)} \right],
 \end{equation}
which is then inserted into the partition function giving
\begin{align}
  Z &= \int\mathfrak{D}[\bar c, c, \bar \phi, \phi, \Sigma]\nonumber\\
     &\quad\exp\left[- S_F[\bar c,c] -\sum_{xy}{\bar\phi_{xy} \phi_{xy} + {1 \over 2}\Sigma_{xy}^{\sigma}\Sigma_{yx}^{\sigma} \over \lambda(x-y)}\right].
\end{align}

In principle, the interaction term in Eq.~\eqref{eq:SF} can be decoupled in three possible channels~\cite{AltlandSimons} -- Cooper, Density, and Exchange -- which capture different physical phenomena. 
The Cooper channel is apposite for describing superconductivity, the density channel encapsulates density fluctuations, and the exchange channel describes electron-hole interactions. 
Weighting these channels is non trivial, and while there are examples of multichannel HS decompositions in the literature,~\cite{Hirsch1983,Ubbens1992,Kleinert2011} we use physical arguments to proceed. 
Since we are interested in describing singlet superconductivity, it is natural to decompose in the Cooper channel the part of the interaction in Eq.~\eqref{eq:SF} with opposite spins. 
Hence, we can identify $\phi$ as being related to fermion pairing and superconductivity.
The term with the same spins is then decomposed in the density channel, since the spins appearing on the fermions are not pertinent, and thus $\Sigma$ acts as a collective density fluctuation.
This decomposition of a single interaction into two different channels, which are treated on equal footing, is one of the main tenets of Eliashberg theory.~\cite{Chubukov2020}
Our neglect of decoupling the interaction in the exchange channel will preclude this analysis from obtaining Kohn-Luttinger corrections, 
which have recently been considered in the diagrammatic framework.~\cite{Phan2020, Chubukov2020}

We now shift the fields according to the transformations:
\begin{align}
  \bar \phi_{xy} &\rightarrow \bar{\phi}_{xy}  - \lambda(x-y) \bar{c}_{x \uparrow} \bar{c}_{y \downarrow},\\
  \phi_{xy} &\rightarrow \phi_{xy} - \lambda(x-y) c_{y \downarrow} c_{x \uparrow},\\
  \Sigma_{xy}^{\sigma} &\rightarrow \Sigma_{xy}^{\sigma} + i\lambda(x-y) \bar{c}_{y\sigma}c_{x \sigma}.
\end{align}
Note that, while $\Sigma(x)$ is a real field, it is perfectly acceptable to shift it by a complex quantity. 
In the case of the integral over a real function of a real variable, such a procedure is tantamount to shifting the contour of integration into the complex plane.
The resulting action now identically cancels the fermionic interaction term at the expense of a coupling to these new Hubbard-Stratonovitch fields $\bar \phi$, $\phi$, and $\Sigma$. 
This transformation results in a new action given by
\begin{align}
  S[\bar{c}, c, \bar{\phi}, \phi, \Sigma] 
    &= \sum_{x,y}\bigg[{\bar\phi_{xy} \phi_{xy} + {1 \over 2}\Sigma_{xy}^{\sigma}\Sigma_{yx}^{\sigma} \over \lambda(x-y)}
      -\bar{c}_{x \sigma} \, G_{0}^{-1}(x-y) \, c_{y \sigma} \nonumber\\
    &\quad - \phi_{xy}\bar{c}_{x\uparrow} \bar{c}_{y \downarrow} - \bar{\phi}_{xy}c_{y \downarrow} c_{x\uparrow} + i \Sigma_{xy}^{\sigma} \bar{c}_{x \sigma}  c_{y \sigma}\bigg].
\end{align}
We consider translation-invariant systems, and thus the HS fields
$\bar\phi, \phi,$ and $\Sigma$ depend only on a relative coordinate. 
After performing the Fourier transform of the action, the momentum-space form is expressed as
\begin{align}
S[\bar{c}, c, \bar{\phi}, \phi, \Sigma] &= \sum_{k,k'}{\bar\phi(k) \phi(k') + {1 \over 2}\Sigma^\sigma(k)\Sigma^\sigma(k') \over \lambda(k-k')}  \nonumber\\
   &\quad - \sum_{k}\bigg[ \bar{c}_{k \sigma}\Big(G_{0}^{-1}(k) - i \Sigma^\sigma(k)\Big)c_{k \sigma} \nonumber\\
   &\quad + \phi(k)\bar{c}_{k\uparrow} \bar{c}_{-k \downarrow} + \bar{\phi}(k) c_{-k \downarrow} c_{k\uparrow}\bigg].
\end{align}

\subsection{Integrating out the fermions}

We now introduce the Nambu fields $\psi_{k}$ and $\bar{\psi}_{k}$, defined by
$\psi_{k} = (c_{k \uparrow},~ \bar{c}_{-k \downarrow})^{T}$ and $\bar{\psi}_{k} = (\bar{c}_{k \uparrow},~ c_{-k \downarrow})$, which facilitate writing the action as
\begin{align}
S[\bar \psi, \psi, \bar\phi, \phi, \Sigma] &= -\sum_{k} \bar{\psi}_k \mathcal{G}^{-1}(k) \psi_{k} \nonumber\\
&\quad + \sum_{k,k'}{ \bar\phi(k) \phi(k') + {1 \over 2}\Sigma^\sigma(k)\Sigma^\sigma(k') \over \lambda(k-k')},
\end{align}
where we define $G_{n,\sigma}^{-1}(k) \equiv G_{0}^{-1}(k) - i\Sigma^\sigma(k)$ and the inverse Nambu Green's function is  
\begin{align}
  \mathcal{G}_{k}^{-1} \equiv
  \begingroup
  \renewcommand*{\arraystretch}{1.5}
  \begin{pmatrix}
    G_{n,\uparrow}^{-1}(k) & \phi(k)\\
    \bar{\phi}(k) & -G_{n,\downarrow}^{-1}(-k)
  \end{pmatrix}.
  \endgroup
\end{align}
The action is now a quadratic form in the Nambu fields $\bar{\psi}$ and $\psi$, and hence the functional integral over these fields can be performed exactly to produce
\begin{align}
 \label{eq:SHST}
&\int\mathfrak{D}[\bar \psi, \psi]
      \exp\left(-S[\bar \psi, \psi, \bar\phi, \phi, \Sigma]\right) \nonumber
    \\&=  \exp\biggl(\mathrm{Tr}\ln\left(-\beta \mathcal{G}^{-1} \right)
  -\sum_{k.k'}{\bar\phi(k) \phi(k') + {1 \over 2}\Sigma^\sigma(k)\Sigma^\sigma(k') \over \lambda(k-k')} \biggr) \nonumber\\
    & \equiv  \exp\left( - S_{\text{HS}}[\bar\phi, \phi, \Sigma] \right).
\end{align}
The trace operation Tr represents a matrix trace over the Nambu indices and an integration over the spatial degrees of freedom. 
The original, microscopic description of electrons interacting with phonons has been reformulated as a description in terms of a Hubbard-Stratonovich action $S_{\text{HS}}$ which depends on the fields $\bar{\phi}, \phi,$ and $\Sigma$. 
Excitations in the field $\phi$ are coupled to the annihilation of two fermions,  whereas the $\Sigma$ field is coupled to fermion density fluctuations.

\section{Hubbard-Stratonovich action analysis}
\label{sec:HST}
\subsection{Saddle-point conditions}

For the field configurations where the effective action is slowly varying with respect to the HS fields, we expect large contributions to the partition function. 
Enforcing that the effective action be stationary with respect to the HS field is known as a saddle-point condition, and it determines the saddle-point, or mean-field, value of $\phi$ and $\Sigma$. 
For the HS action in Eq.~\eqref{eq:SHST}, the saddle-point condition for $\Sigma$ is
\begin{align}
  0 &= {\delta S_{\text{HS}}[\bar\phi,~ \phi,~ \Sigma] \over \delta \Sigma_k^\sigma} \nonumber\\
     &= \sum_{k'} \lambda^{-1}({k-k'}) \Sigma^\sigma(k') + i \delta_{k,k'}\sum_{k'} \mathrm{Tr}\Bigg[
      \mathcal{G}_{k'}
      \begin{pmatrix}
        \delta_{\sigma\uparrow} & 0 \nonumber\\
        0 & -\delta_{\sigma\downarrow}
      \end{pmatrix}   \Bigg] \nonumber\\
    &= \sum_{k'} \lambda^{-1}(k-k') \Sigma^\sigma(k') + i {G_{n\bar\sigma}^{-1}(-k) \over G_{n,\sigma}^{-1}(k)G_{n, \bar\sigma}^{-1}(-k) + |\phi(k)|^2}. 
\end{align}

Using the identity $\delta_{k,k'} = \sum_{k''} \lambda(k-k'')\lambda^{-1}(k''-k')$, we can simplify the above equation to solve for $\Sigma^{\sigma}$. 
By taking advantage of the spin symmetry of the system, i.e., $\mu_\uparrow=\mu_\downarrow$ and no magnetic field is present, we can define $ \Sigma \equiv i\Sigma^\uparrow = i\Sigma^\downarrow$ and
$ G_n \equiv G_{n\uparrow} = G_{n\downarrow}$, to obtain
\begin{equation}
\label{eq:Sigmamf}
\Sigma(k) = \sum_{k'}\lambda(k-k'){G_{n}^{-1}(-k) \over G_{n}^{-1}(k)G_{n}^{-1}(-k) + |\phi(k)|^2}.
\end{equation}
From this equation we now recognize that $G_{n}$ is a Green's function that is dressed by $\Sigma$. 
That is, the mean-field value of the field $\Sigma^{\sigma}$ plays the role of a self-energy.
In the conventional~\cite{Rickayzen,Bennemann,Marsiglio2020} formulation of Eliashberg theory, based on the equation of motion technique, it is not apparent that the self energy $\Sigma$ arises from the saddle-point value of an HS field. 
The importance of this result, which naturally appears in the functional-integral approach, is that it enables one to consider fluctuations beyond the saddle-point condition and possible corrections to the mean-field Eliashberg equations.
This will be explored further in the next subsection. 

The saddle-point analysis of the pairing field $\phi$ can also be performed, leading to
\begin{equation}
  0 = {\delta S_{\text{HS}}[\bar\phi ~ \phi~ \Sigma] \over \delta \bar{\phi}_k} = \sum_{k'} \lambda_{k-k'}^{-1} \phi(k') 
  + \mathrm{Tr}\Bigg[ \mathcal{G}_{k}
     \begin{pmatrix}
        0 & 0\\
        1 & 0
      \end{pmatrix}\Bigg].
\end{equation}
Solving this equation for $\phi$ gives
\begin{align*}
  \phi(k) = \sum_{k'} \lambda({k-k'}) {\phi(k') \over  G_{n}^{-1}(k')G_{n}^{-1}(-k')+ |\phi(k')|^2}~~. \numberthis
\end{align*}
This homogeneous equation has the familiar form of a gap equation for the superconducting order parameter $\phi$. 
One possible solution to the above equation is $\phi=0$, which represents a normal-state system.
For this case, $\Sigma$ corresponds to a self energy arising solely from normal-state electron-phonon interactions. 
Similarly, $G_{n}$ would then be a dressed normal-state Green's function. For non-zero $\phi$, however, $G_{n}$ is modified by superconducting interactions. 
The term multiplying the electron-phonon interaction $\lambda$ in the sum in Eq.~\eqref{eq:Sigmamf} can be identified as a dressing of $G_n$ due to an anomalous self energy arising from $|\phi|^2$ which encodes the pairing of fermions. 
Hence, we denote the full Green's function by
\begin{equation}
 G(k) \equiv {G_{n}^{-1}(-k) \over G_{n}^{-1}(k)G_{n}^{-1}(-k) + |\phi(k)|^2}.
\end{equation}
The self-consistent equation for the self energy is then
\begin{equation}
\Sigma(k) = \sum_{k'}\lambda(k-k') G(k').
\end{equation}

\begin{figure}[h]
    \centering 
    \includegraphics[width=0.8\linewidth]{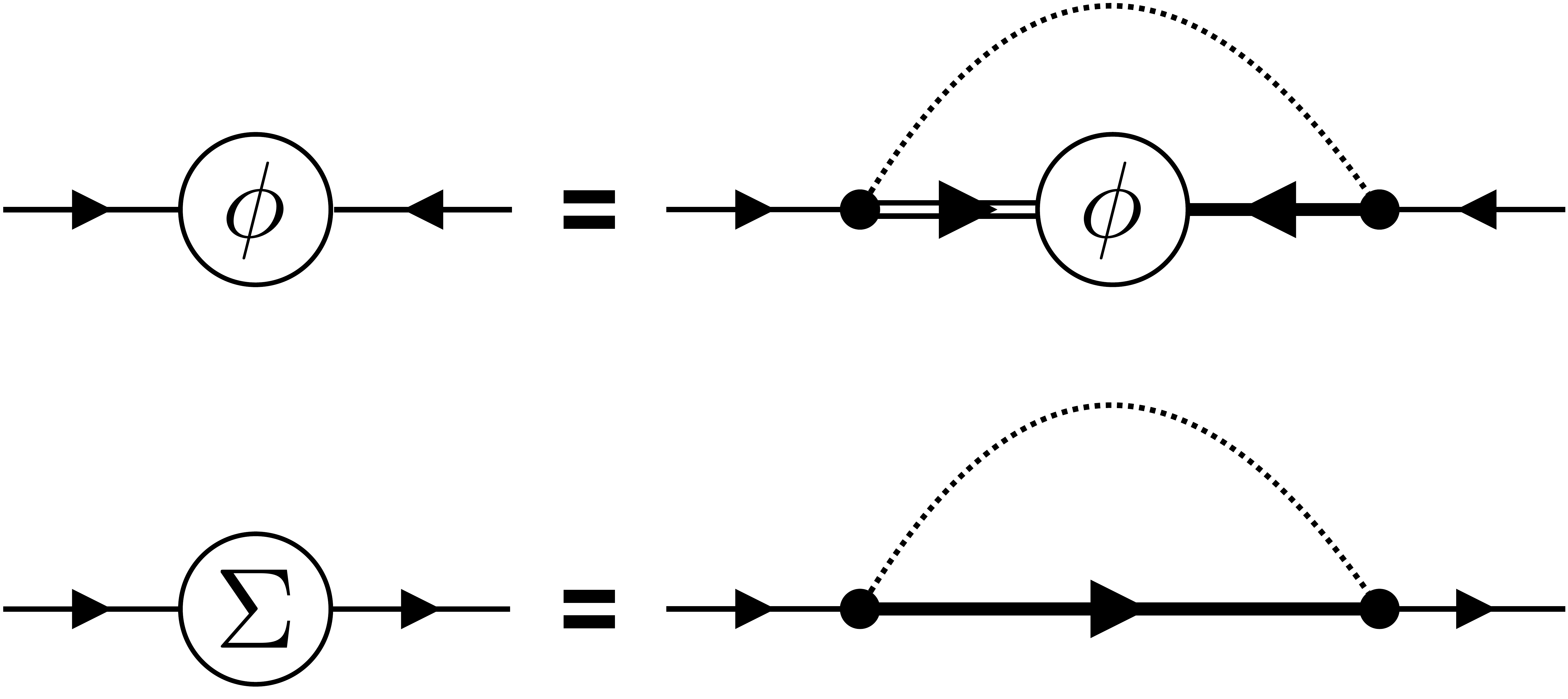} 
    \caption{A diagrammatic representation of the mean-field Eliashberg equations \eqref{eq:phieq} and \eqref{eq:sigmaeq}. The dotted lines terminated by circles denote $\lambda$, i.e., the phonon propagator and the coupling constants. 
      The double-struck solid lines denote $G_n$ and the bold solid lines are the full fermion propagator $G$. The lines at the ends of the diagram are external legs.}\label{fig:mfeq}
\end{figure}

Together, these mean-field equations produce the standard~\cite{Rickayzen,Bennemann,Marsiglio2020} Eliashberg equations depicted in Fig.~\ref{fig:mfeq}:
\begin{align}
\label{eq:phieq}\phi(\B k, i\omega_n) &= \sum_{\B k', i\omega_m} \lambda_{\B{k-k'}}(i\omega_n - i\omega_m) F(\B k', i\omega_m). \\
\label{eq:sigmaeq}\Sigma(\B k, i\omega_n) &= \sum_{\B k', i\omega_m} \lambda_{\B{k-k'}}(i\omega_n - i\omega_m) G(\B k', i\omega_m). \\
\label{eq:gneq}G_n(\B k, i\omega_n) &= {1 \over G_0^{-1}(\B k, i\omega_n) - \Sigma(\B k, i\omega_n)}. \\
\label{eq:geq}G(\B k, i\omega_n) &= {G_{n}^{-1}(-\B k, -i\omega_n) \over G_{n}^{-1}(\B k, i\omega_n)G_{n}^{-1}(-\B k, -i\omega_n) + |\phi(\B k, i\omega_n)|^2}. \\
\label{eq:feq}F(\B k, i\omega_n) &= {\phi(\B k, i\omega_n) \over G_{n}^{-1}(\B k, i\omega_n)G_{n}^{-1}(-\B k, -i\omega_n) + |\phi(\B k, i\omega_n)|^2}. 
\end{align}

\begin{figure}[h]
    \centering 
    \includegraphics[width=0.9\linewidth]{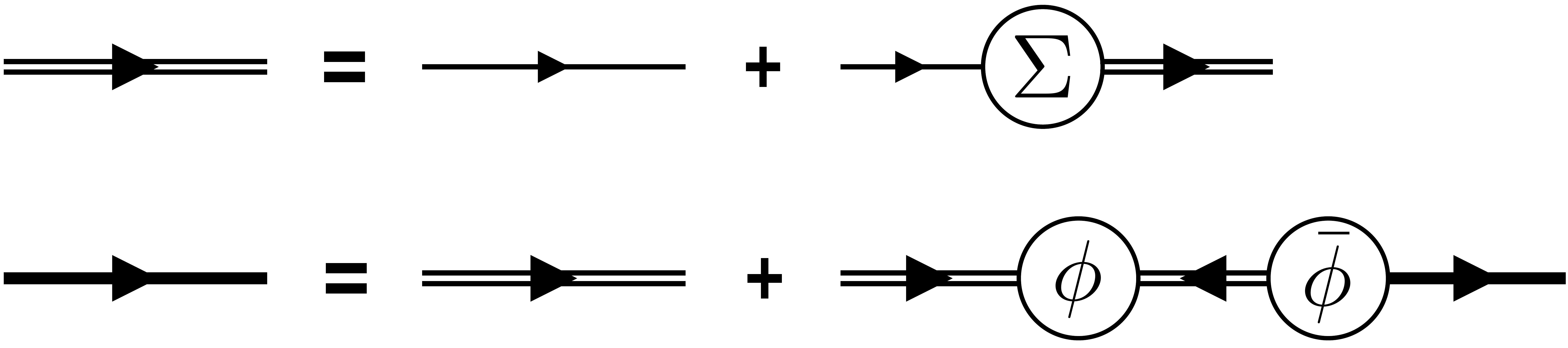} 
    \caption{A diagrammatic representation of the dressed propagators appearing in the Eliashberg equations~\eqref{eq:gneq} and \eqref{eq:geq}}
    \label{fig:dressedprops}
\end{figure}

\subsection{Gaussian fluctuations in the normal state}

In this section we shall focus on normal-state fluctuations. We allow $\phi$ and $\Sigma$ to depend on both a \emph{relative} and a \emph{centre of mass} coordinate:
\begin{equation}
  \phi_{xy} = \phi\left(x-y;~ {x+y \over 2} \right); \Sigma^\sigma_{xy} = \Sigma^\sigma\left(x-y;~ {x+y \over 2} \right).
\end{equation}
As a result, the Fourier transforms of $\phi_{xy}= \bar{c}_{x\uparrow}\bar{c}_{y\downarrow}$ and $\Sigma_{xy}^{\sigma} = \bar{c}_{x\sigma}c_{y\sigma}$ are 
$\phi(k; q) = \bar{c}_{k+q/2 \sigma}\bar{c}_{-k+q/2 \sigma}$ and $\Sigma^\sigma(k; q) = \bar{c}_{k + q/2 \sigma}c_{k - q/2 \sigma}$ respectively.
We restrict our attention to small fluctuations away from the saddle-point condition, where we enforce that $q=0$, and consider
\begin{align}
  \phi(k; q) &\equiv \phi(k)\delta_{q,0} + \eta(k; q)~~,\\
  \Sigma^\sigma(k; q) &\equiv \Sigma^\sigma(k)\delta_{q,0} + \theta^\sigma(k; q).
\end{align}
Here, $\phi(k)$ and $\Sigma(k)$ are given by the mean-field solutions to the Eliashberg equations~\eqref{eq:phieq}-\eqref{eq:feq} and $\eta(k,q)$ and $\theta(k, q)$ are small quantities that enable the action to be expanded to quadratic order in $\eta$ and $\theta$. 
We emphasize that the momentum $k$ in $\eta(k, q)$ is not transferred but rather merely acts as a label, whereas the momentum $q$ is transferred. 
This is required on the grounds that $k$ is a fermionic Matsubara frequency whereas $\phi$ (and hence $\eta$) are bosonic fields.

We now expand the inverse Green's function about its mean-field value via $\mathcal{G}^{-1} = \mathcal{G}_{\mathrm{mf}}^{-1} \Big(1 -\mathcal{G}_{\mathrm{mf}} \Lambda_\eta - \mathcal{G}_{\mathrm{mf}}\Lambda_\theta\Big)$  where
\begin{align}
  \Lambda_\eta &= \begin{pmatrix}
    0               & -\eta(k; q)\\
    -\bar\eta(k; -q) & 0
  \end{pmatrix}, \\
  \Lambda_\theta(k; q) &\equiv \begin{pmatrix}
    i\theta(k;q)      & 0\\
    0                 & -i\theta(-k; q) 
  \end{pmatrix}.
\end{align}
Hence, the action may be written as
\begin{align}
&S_{\text{HS}}[\bar{\phi}, \phi, \bar{\eta}, \eta, \Sigma, \theta] \nonumber\\
  & = \sum_{xy} {(\bar{\phi}_{xy} + \bar\eta_{xy})(\phi_{xy} + \eta_{xy})
    + {1 \over 2}(\Sigma^\sigma_{xy} + \theta^\sigma_{xy})(\Sigma^\sigma_{yx} + \theta^\sigma_{yx}) \over \lambda(x-y)} \nonumber\\
   & \quad - \mathrm{Tr}\ln\Big[\mathcal{G}_{\mathrm{mf}}^{-1} (1 - \mathcal{G}_{\mathrm{mf}}\Lambda_{\eta} - \mathcal{G}_{\mathrm{mf}}\Lambda_{\theta}) \Big].
\end{align}
Using the fact that the perturbations $\eta$ and $\theta$ are small, we can expand the logarithm to quadratic order, noting that the saddle-point conditions ensure that the terms linear in $\phi$ and $\Sigma$ vanish identically. 
The HS action is thus
\begin{align}
S_{\text{HS}}[\bar{\phi}, \phi, \bar{\eta}, \eta, \Sigma,\theta] 
& = S_{\mathrm{mf}}[\bar{\phi}, \phi, \Sigma] + \sum_{x,y}{\bar{\eta}_{xy}\eta_{xy} + {1\over 2}\theta^\sigma_{xy}\theta^\sigma_{yx}  \over \lambda(x-y)} \nonumber\\
&\quad +{1 \over 2}\mathrm{Tr}\Big[(\mathcal{G}_{\mathrm{mf}}\Lambda_\eta +  \mathcal{G}_{\mathrm{mf}}\Lambda_\theta)^2 \Big].\label{eq:SHS}
\end{align}

\begin{figure}[t]
    \centering 
    \includegraphics[width=0.9\linewidth]{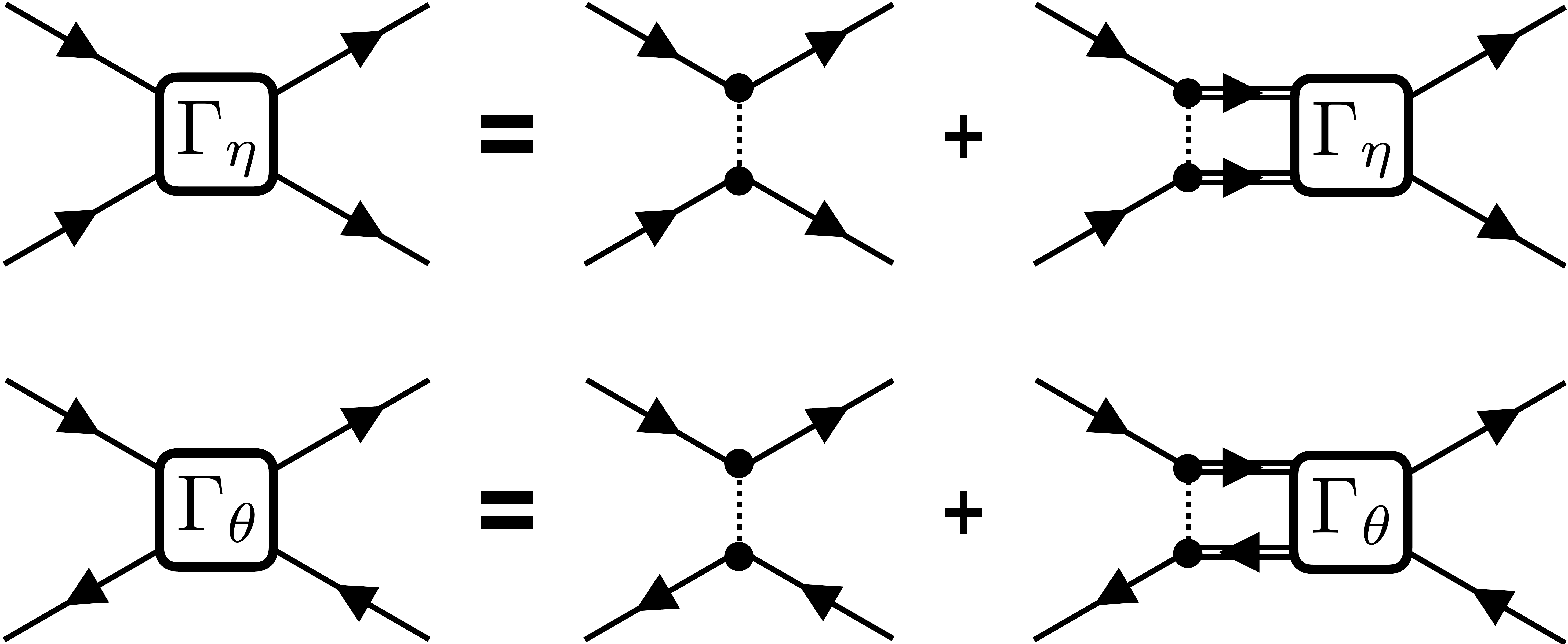} 
    \caption{A diagrammatic representation of the Dyson equation for the normal-state Eliashberg fluctuation propagators.}
    \label{fig:dressedprops}
\end{figure}

We assume that the system under study is above the superconducting critical temperature $T_{c}$ and set $\phi=\bar\phi=0$. 
After performing some matrix algebra, the squared trace term simplifies to the following result expressed in momentum space
\begin{align}
  {1 \over 2}&\mathrm{Tr}\Big[(\mathcal{G}_{\mathrm{mf}}\Lambda_\eta +  \mathcal{G}_{\mathrm{mf}}\Lambda_\theta)^2 \Big] \nonumber\\
  &=  {1 \over 2}\mathrm{Tr}\Big[\mathcal{G}_{n}\Lambda_\eta \mathcal{G}_{n}\Lambda_\eta \Big]
    + {1 \over 2}\mathrm{Tr}\Big[\mathcal{G}_{n}\Lambda_\theta \mathcal{G}_{n}\Lambda_\theta \Big] \nonumber\\
  &= -\sum_{k, q} \Big[ \theta^\sigma(k+q/2;q)G_{n}(k)G_n(k-q)\theta^\sigma(k+q/2;-q) \nonumber\\
  & \quad + \bar\eta(k+q/2;q)G_{n}(k)G_{n}(q-k)\eta(k+q/2;q) \Big].
\end{align}
Therefore, the HS action is now in the form $S_{\text{HS}}[\bar{\phi}, \phi, \Sigma, \bar{\eta}, \eta, \theta] =S_{\text{mf}}[\bar{\phi}, \phi, \Sigma] +S[\bar{\eta},\eta,\theta]$, where 
\begin{align}
\label{eq:S_fluc}
  &S[\bar{\eta}, \eta, \theta]\\
  &=\sum_{k,k',q}\Bigg[\bar\eta(k+q/2;q)~ \Gamma^{-1}_\eta(k,k';q)~ \eta(k'+q/2;q) \nonumber\\
  &~\quad + {1 \over 2}\theta^\sigma(k+q/2;q)~ \Gamma^{-1}_\theta(k,k';q)~ \theta^\sigma(k'+q/2;-q) \Bigg].
\end{align}
We have introduced $\Gamma_\eta(k,k'; q)$ and $\Gamma_\theta(k,k';q)$ as fluctuation propagators for the fields $\eta$ and $\theta$ respectively, and they are defined by the following Dyson equations
\begin{align}
\label{eq:GammaEta}  \Gamma_{\eta}^{-1}(k, k'; q)  &=\lambda^{-1}(k-k') - \delta(k-k') G_n(k)G_n(q-k), \\
\label{eq:GammaTheta}  \Gamma_{\theta}^{-1}(k, k'; q) &= \lambda^{-1}(k-k')  - 2\delta(k-k') G_n(k)G_n(k-q).
\end{align}
The Dyson equations for the fluctuation propagators are shown in Fig.~\ref{fig:dressedprops}. Notice that $\Gamma_{\eta}$ has a particle-particle susceptibility whereas $\Gamma_{\theta}$ has a particle-hole susceptibility.
After performing the functional integral over the fields $\bar{\eta}, \eta$, and $\theta$ we obtain the fluctuation action $S_{\text{fluc}}$, defined by
$\int\mathfrak{D}[\bar{\eta},\eta,\theta]\exp\left(-S[\bar{\eta},\eta,\theta]\right)  = \exp(-S_{\text{fluc}})$, where 
\begin{equation}
S_{\text{fluc}} = \mathrm{Tr}\ln\left[-\Gamma^{-1}_{\eta} \right] + \frac{1}{2}\mathrm{Tr}\ln\left[-\Gamma^{-1}_{\theta} \right].\label{eq:Sfluc}
\end{equation}
The function $\Gamma_\eta(k,k';q)$ represents the propagator for two electrons above $T_c$, with relative momentum $k$ and centre of mass momentum $q$, to form a short-lived Cooper pair before separating into two electrons with relative momentum $k'$. 
The condition that $\Gamma^{-1}_{\eta}(k,k';0) \rightarrow 0$ would imply that these fluctuations proliferate at $q=0$, signalling a phase transition to the superconducting state.

We now compare the superconducting fluctuation propagator derived here with that found in Eq.~(1) of Ref.~\onlinecite{Narozhny1993}, which in our notation reads
\begin{align}
\label{eq:narozhnyGamma}
& \Gamma_\eta(p,p';q) = \lambda(p-p') \nonumber\\ &+\sum_{p''}\lambda(p-p'')G_n(p'')G_n(q-p'')\Gamma_\eta(p'',p';q).
\end{align}
Here we have taken the liberty of restoring the coupling constants $g_{\B q}$ in the expression in Ref.~\onlinecite{Narozhny1993} and replaced their $\mathcal{D}$ expressions by $\lambda$.  
If we multiply the above equation by $\lambda^{-1}(k-p)\Gamma^{-1}_{\eta}(p',k';q)$ and perform the integral over $p$ and $p'$, then we obtain 
\begin{align}
&\sum_{p,p'}\lambda^{-1}(k-p)\Gamma_\eta(p,p';q)\Gamma^{-1}_{\eta}(p',k';q) \nonumber\\
& = \sum_{p,p'}\lambda^{-1}(k-p)\lambda(p-p')\Gamma^{-1}_{\eta}(p',k';q) \nonumber\\
&\quad + \sum_{p,p',p''}\lambda^{-1}(k-p)\lambda(p-p'')\Gamma^{-1}_{\eta}(p',k';q)\nonumber\\&\quad\times G_n(p'')G_n(q-p'')\Gamma_\eta(p'',p';q).
\end{align}
Performing the remaining momentum integrals results in
\begin{align}
&\lambda^{-1}(k-k')= \sum_{p'} \delta(k-p') \Gamma^{-1}_{\eta}(p',k';q) \nonumber\\ &\quad + \sum_{p',p''}\delta(k-p'')\Gamma_\eta(p'',p';q)\Gamma^{-1}_{\eta}(p',k';q)G_n(p'')G_n(q-p'') \nonumber\\
&\hspace{15mm} = \Gamma^{-1}_{\eta}(k,k';q) + \delta(k-k')G_n(k)G_n(q-k).
\end{align}
Rearranging this expression then reproduces the result we obtained in Eq.~\eqref{eq:GammaEta} through functional-integral methods. 
We emphasize that, in addition to $\Gamma_\eta$, we have also derived the propagator $\Gamma_\theta$, which parameterizes density fluctuations in the $\Sigma$ field.
Interestingly, the Dyson equation for $\Gamma_\theta$ has a similar form to the phonon propagator in Migdal-Eliashberg theory,~\cite{Marsiglio1990,Marsiglio1991,Bennemann} where the bare phonon propagator is corrected by a phonon self energy involving a loop of two fermionic Green's functions. Nevertheless, note that, in the normal-state under consideration, the Green's function is fixed as $G_{n}$ and the (mean-field) self energy is defined in terms of $G_{n}$ and $\lambda$, that is, it is not $\Gamma_{\theta}$ which appears in the self-consistency conditions.

\section{Fluctuation response}
\label{sec:FlucRes}

\begin{figure}[h]
    \centering 
    \includegraphics[width=0.8\linewidth]{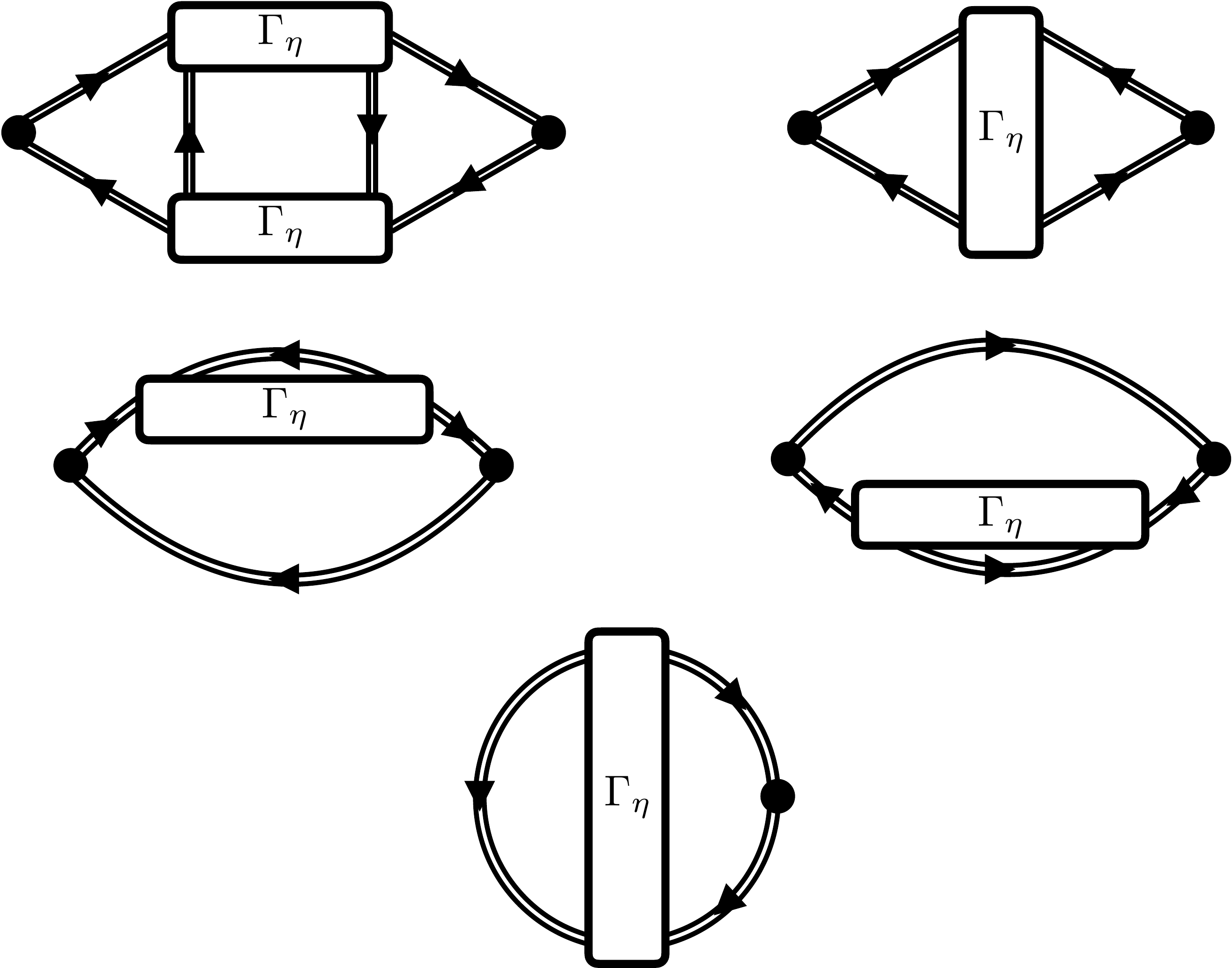} 
    \caption{Cooper-channel fluctuation response diagrams in the normal state. The top left diagram is the AL diagram, next is the MT diagram, followed by the two DOS diagrams, and finally the diamagnetic diagram.}
    \label{fig:FluctuationResponse1}
\end{figure}
\begin{figure}[h]
    \centering 
    \includegraphics[width=0.8\linewidth]{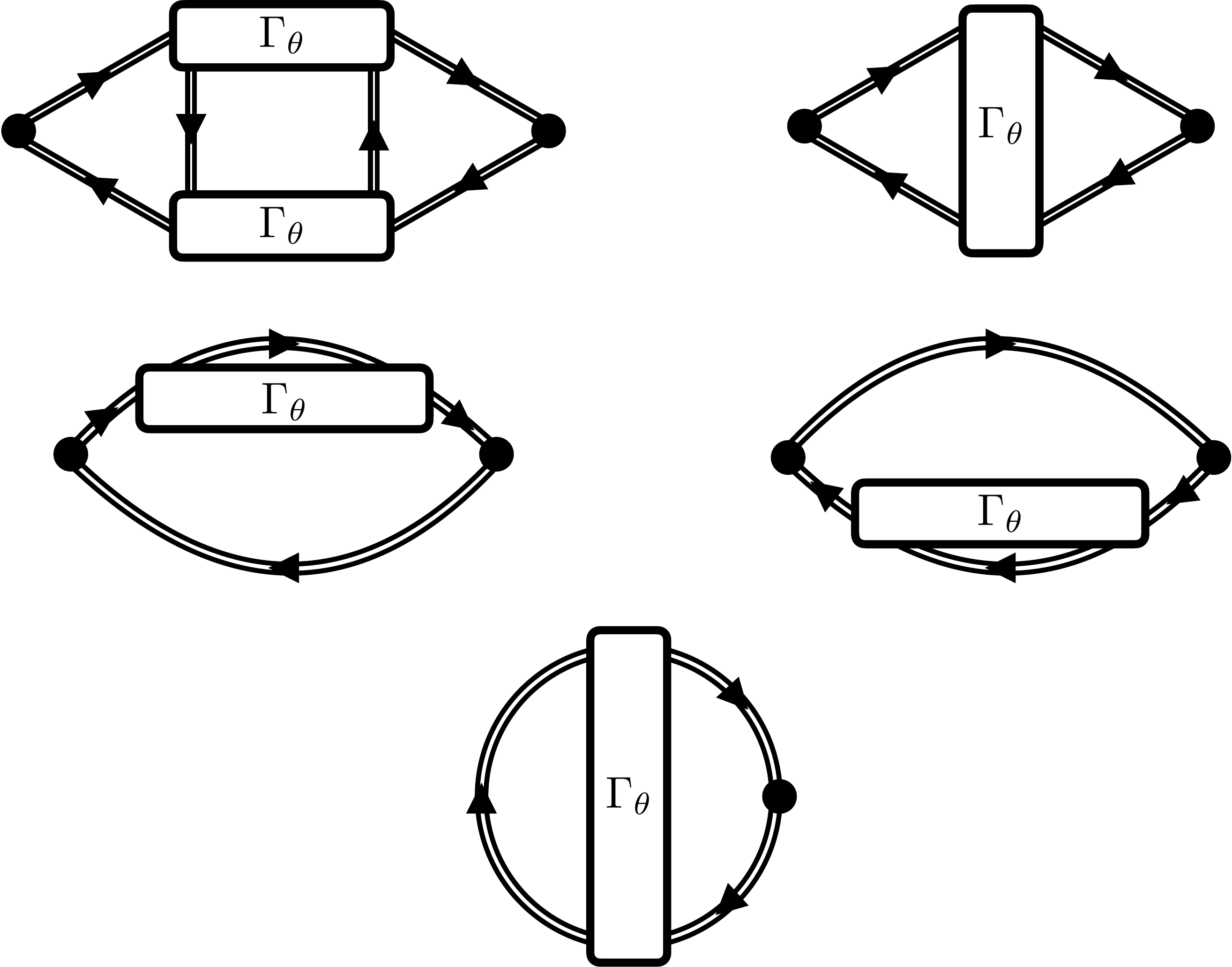} 
    \caption{Density-channel fluctuation response diagrams in the normal state.}
    \label{fig:FluctuationResponse2}
\end{figure}

In this section we derive the normal-state fluctuation response based on the propagators derived in the previous section. 
To derive the electromagnetic response, the standard method~\cite{AltlandSimons} is to introduce an external background gauge field $A_{\mu}$ and apply a minimal coupling procedure. 
To simplify the derivation, we assume that $p,p'\approx p_{F}$ in the propagators $\Gamma_{\eta}(p,p';q)$ and $\Gamma_{\theta}(p,p';q)$, where $p_{F}$ is the Fermi momentum. 
Thus, if we let $q$ denote the momentum of the external gauge field $A_{\mu}$, then we can derive the response in the same manner as the conventional fluctuation response.~\cite{SvidzinskiiBook,Boyack2018}

Using the results from Eq.~\eqref{eq:SHS} and Eq.~\eqref{eq:Sfluc}, the full normal-state action is given by 
\begin{align}
S(T>T_{c}) &= -2\sum_{p,p'}\Sigma(p)\lambda^{-1}(p-p')\Sigma(p') -2\sum_{p}\ln\left[-G_{n}^{-1}\right] \nonumber\\ 
&\quad + \frac{1}{2}\sum_{p}\ln\left[-\Gamma^{-1}_{\theta}(p) \right] + \sum_{p}\ln\left[-\Gamma^{-1}_{\eta}(p) \right] \nonumber\\
& = -2\sum_{p}\left\{\ln\left[\Sigma(p)-G_{0}^{-1}(p)\right] +\Sigma(p)G_{n}(p)\right\}  \nonumber\\ 
&\quad  + \frac{1}{2}\sum_{q}\ln\left[-\Gamma^{-1}_{\theta}(q) \right] + \sum_{q}\ln\left[-\Gamma^{-1}_{\eta}(q) \right].
\end{align}
In the last step we used Eq.~\eqref{eq:Sigmamf} to simplify the first term. 
This result agrees with Eq.~(5) of Ref.~\onlinecite{Narozhny1994}, which considered first-order perturbation theory in the electron-phonon interaction and suitably modified Eq.~(1) of Ref.~\onlinecite{Eliashberg1963}. 
We emphasize that the result above has fixed $\Sigma$ at its mean-field value, but incorporated fluctuations about this quantity that give rise to $S_{\text{fluc}}$.
As a result, this action is not extremal with respect to variations in $\Sigma$, and therefore it does not exactly reduce to the result in Ref.~\onlinecite{Eliashberg1963}.
Rather, we have presented here a rigorous derivation of the result suggested in Ref.~\onlinecite{Narozhny1994}.

The fluctuation action, in the presence of the external field, has the form $S_{\text{fluc}}[A]=\text{Tr} \ln\left[ -\Gamma^{-1}_{\eta}[A]\right]+\frac{1}{2}\text{Tr} \ln\left[- \Gamma^{-1}_{\theta}[A]\right]$.
The response is then obtained using the definition $K^{\mu\nu}=\left.\delta^{2}S[A]/\delta A_{\mu}\delta A_{\nu}\right|_{A=0}$. Performing these derivatives, we obtain 
\begin{equation}
  K^{\mu\nu}_{\text{fluc}}=\sum_{a=\eta,\theta}\left\{-\text{Tr} \Gamma_{a}\frac{\delta \Gamma^{-1}_{a}}{\delta A_{\mu}}\Gamma_{a}\frac{\delta \Gamma^{-1}_{a}}{\delta A_{\mu}}
+\text{Tr}\Gamma_{a}\frac{\delta^{2}\Gamma^{-1}_{a}}{\delta A_{\mu}\delta A_{\nu}}\right\}.
\end{equation}
The first term is the Aslamazov-Larkin (AL) diagrams, one with the fluctuation propagator $\Gamma_{\eta}$ describing the transport of Cooper pairs and the other with the fluctuation propagator $\Gamma_{\theta}$ describing the transport of phonon fluctuations. The second term comprises the Maki-Thompson (MT), Density of States (DOS), and diamagnetic contributions.~\cite{VarlamovBook,SvidzinskiiBook,Boyack2018} 
These diagrams are shown in Figs.~\ref{fig:FluctuationResponse1}--\ref{fig:FluctuationResponse2}.

Here we compute the singular contribution to the diamagnetic susceptibility for the Eliashberg fluctuation response. 
The Kubo formula for diamagnetic susceptibility is:~\cite{Vignale1988}
\begin{equation}
\chi_{\text{dia}}=-\frac{1}{q^{2}}\left.K^{xx}\left(i\Omega_{m}=0,q^{y}\right)\right|_{q^{y}\rightarrow0,q^{x}=q^{z}=0}.
\end{equation}
The absence of a normal-state Meissner effect requires~\cite{Rickayzen} that $K^{xx}(i\Omega_{m}=0,\mathbf{q}\rightarrow0)=0$, thus the above Kubo formula is well defined. 
Near the transition temperature the fluctuation propagator becomes singular, and since the AL diagram has one more propagator than the other fluctuation diagrams, we may expect this term to provide the sole contribution to the diamagnetic susceptibility.~\cite{Aslamazov1975}
Since $\chi$ is a thermodynamic quantity, no analytic continuation needs to be performed, and thus there is no anomalous Maki-Thompson contribution to consider.~\cite{VarlamovBook} 
As a result, the AL term does indeed provide the dominant singular contribution near $T_{c}$. 

The vertex function $\Gamma_{\eta}$ can be written as~\cite{Narozhny1993}: $\Gamma_{\eta}(p,p';q) = \phi(p)L(q)\bar{\phi}(p')$, 
where $\phi$ is the solution to the linearized version of Eq.~\eqref{eq:phieq} and $L$ is the fluctuation propagator which depends only on the centre of mass momentum $q$. 
Following Refs.~\onlinecite{Narozhny1993,VarlamovBook}, the AL diagram for the fluctuation Cooper pairs, i.e., $\eta$ fluctuations, in the top left diagram in Fig.~\ref{fig:FluctuationResponse1}, is then
\begin{align}
K_{\text{AL},\eta}^{xx}\left(0,q^{y}\right) &= -\frac{4e^{2}}{m^{2}}T\sum_{\B p, \epsilon_{m}}L\left(i\epsilon_{m},{\bf p}_{+}\right)C^{x}\left(p_{+},p_{-}\right)\nonumber\\
&\hspace{22mm}\times L\left(i\epsilon_{m},{\bf p}_{-}\right)C^{x}\left(p_{-},p_{+}\right).
\end{align}
Here $\mathbf{p}_{\pm}=\mathbf{p}\pm\mathbf{q}/2$, and $L$ is the fluctuation propagator introduced above. 
The vertex $C^{x}$ is defined in position space by $C^{x}(x,y,z)\sim\delta L^{-1}[A](x,y)/\delta A^{x}(z)$; for the momentum-space form see Ref.~\onlinecite{Narozhny1993}. The critical regime is characterised by $L^{-1}\left(0,{\bf q}\rightarrow0\right)\rightarrow0$, thus the Matsubara frequency equal to zero provides the dominant contribution and so we set $\epsilon_{m}=0$. Performing the Taylor expansion of the fluctuation propagators and using the Kubo formula, we obtain
\begin{equation}
\chi_{\text{fluc}}=\frac{e^{2}}{m^{2}}T\sum_{{\bf p}}\left[C^{x}\left(p,p\right)\right]^{2}\left(LL^{\prime\prime}-\left(L^{\prime}\right)^{2}\right).
\end{equation}
Here the prime denotes differentiation with respect to $p^{y}$. The vertex $C^{x}\left(p,p\right)$ has the form $C^{x}\left(p,p\right)=p^{x}C\left(0\right)$. 
Performing the derivatives and using integration by parts, we obtain
\begin{equation}
\chi_{\text{fluc}} = -\frac{2}{3}\frac{e^{2}}{m^{2}}T\sum_{{\bf p}}\left[C^{x}\left(p,p\right)\right]^{2}L^{3}(0,\mathbf{p})\left(L^{-1}(0,\mathbf{p})\right)^{\prime\prime}. \label{eq:Chi}
\end{equation}
The inverse retarded fluctuation propagator has the form:~\cite{Narozhny1993}
$L_{\text{R}}^{-1}\left(\Omega,{\bf p}\right)=N\left(\frac{i\Omega}{T_{c}}b-Dp^{2}-\frac{T-T_{c}}{T_{c}}\right).$
Here, $N$ is the single-spin density of states at the Fermi surface, $b$ is a constant, and $D$ is related to the square of the coherence length.~\footnote{This is the analogue of the conventional fluctuation propagator given in Eq.~(6.17) of Ref.~[\onlinecite{VarlamovBook}]; 
in that case $b=\pi/8$, $D=\xi_{3\text{d}}^2$, where the three-dimensional coherence length for a clean system is $\xi_{3\text{d}}=\sqrt{7\zeta(3)v_{F}^2/(48\pi^2T^2)}$.} 
Inserting the propagator into Eq.~\eqref{eq:Chi} and performing the integration, the diamagnetic susceptibility is found to be 
\begin{eqnarray}
\chi_{\text{fluc}} & = & \frac{4}{3}ND\frac{e^{2}}{m^{2}}T\sum_{{\bf p}}\left[p^{x}C\left(0\right)\right]^{2}L^{3}(0,\mathbf{p}) \nonumber\\
 & = & \frac{2}{9\pi^{2}}ND\left[C\left(0\right)\right]^{2}\frac{e^{2}}{m^{2}}T\int_{0}^{\infty}dpp^{4}L^{3}\nonumber \\
 & = & -\frac{T\left[C\left(0\right)\right]^{2}}{24\pi N^{2}D^{\frac{3}{2}}}\frac{e^{2}}{m^{2}}\sqrt{\frac{T_{c}}{T-T_{c}}}.\label{eq:Chi}
\end{eqnarray}
The power-law divergence of the diamagnetic susceptibility is the same as that in Ref.~\onlinecite{Aslamazov1975} for conventional fluctuations. 
However, the presence of the electron-phonon interaction can significantly modify the prefactor appearing above.

In the strong-coupling limit, Ref.~\onlinecite{Narozhny1993} asserts that 
\begin{equation}
N\sim mp_{F},\ b\sim\frac{1}{\lambda},\ D\sim\frac{1}{\lambda^{3}}\frac{v_{F}^{2}}{T_{c}^{2}},\ C\left(0\right)\sim\frac{1}{\lambda^{2}}\frac{p_{F}^{3}}{T_{c}^{2}}.
\end{equation}
The normal-state diamagnetic susceptibility for a spin-$\frac{1}{2}$ electron gas (Landau diamagnetism of a free-electron gas) is~\cite{Ziman} $\chi_{\text{Landau}}=-e^{2}v_{F}/(12\pi^2)$. 
Inserting these results into Eq.~\eqref{eq:Chi}, the ratio of the strong-coupling limit of the fluctuation diamagnetic susceptibility to the normal-state value is found to be
\begin{equation}
\frac{\chi_{\text{fluc}}}{\chi_{\text{Landau}}}\sim\lambda^{\frac{1}{2}}\frac{\pi}{2}\sqrt{\frac{T_{c}}{T-T_{c}}},\quad\lambda\rightarrow\infty.
\end{equation}
In the limit of large electron-phonon coupling, the fluctuation susceptibility is increasingly large due to the prefactor $\lambda^\frac{1}{2}$. Interestingly, the critical temperature also has the same strong-coupling dependence~\cite{Allen1982}: 
$T_{c}/\omega_{E}\sim\lambda^{\frac{1}{2}}$ as $\lambda\rightarrow\infty$, where $\omega_{E}$ is the Einstein frequency. The fluctuation electrical conductivity was also found to have a large prefactor in the limit of large electron-phonon interactions~\cite{Narozhny1993}, albeit with a different power law dependence on $\lambda$.
Thus, in comparison to the seminal results of Ref.~\onlinecite{Aslamazov1975}, when the electron-phonon coupling becomes increasingly large the fluctuation regime for large diamagnetic susceptibility is broadened.

\section{Conclusion}
\label{sec:Conc}

The Eliashberg theory of superconductivity incorporates the dynamical nature of the electron-phonon interaction, and conventionally it is derived within an equation of motion approach. Here we have shown how to derive the mean-field Eliashberg equations using a functional-integral method. Importantly, this analysis illustrates that the electron self energy appears as a Hubbard-Stratonovich field, which enables its fluctuations about the mean-field to be considered. We reproduced the standard mean-field equations as saddle-point conditions of the Hubbard-Stratonovich action, without recourse to intricate diagrammatic arguments. In addition, we considered the Gaussian-fluctuation response and obtained both Cooper- and density-channel excitations. The former had been studied previously, but our systematic analysis naturally obtains both fluctuations at the same time. As a result, we obtained the fluctuation diagrams for Eliashberg theory and have provided the complementary account for what has been done for BCS theory. Furthermore, we computed the fluctuation diamagnetic susceptibility near the critical temperature and determined its strong-coupling form. Our functional approach enables clear pathways for going beyond the traditional Eliashberg theory framework by considering alternative Hubbard Stratonovich  decompositions.

\acknowledgments
This work was supported in part by the Natural Sciences and Engineering Research Council of Canada (NSERC), and by an MIF from the
Province of Alberta. R.B. acknowledges support from the Department of Physics and the Theoretical Physics Institute at the University of Alberta.
\vspace{1.95cm}
~

\bibliographystyle{apsrev4-1}
%

\end{document}